\documentclass[a4paper]{article}

\usepackage{amsmath,amssymb,amsfonts,amsthm,amscd,makeidx,stackrel,stmaryrd,float,epsfig,multicol}
\usepackage[left=3cm,top=2.5cm,bottom=2.5cm,right=3cm]{geometry}
\usepackage{amsthm}
\usepackage{array}
\usepackage{bm}
\usepackage{eucal}
\numberwithin {equation}
{section}
\theoremstyle{definition}
\newcounter{dummy} \numberwithin{dummy}{section}

\newtheorem{oss}[dummy]{Remark}

\numberwithin{equation}{section}
\usepackage{graphicx}
\title{Unsteady rotating laminar flow: analytical solution of Navier-Stokes equations}
\author{Alessio Bocci, Giovanni Mingari Scarpello, Daniele Ritelli}
\date{}

\begin{document}
\maketitle
\begin{abstract}

We provide a integration of Navier-Stokes equations  concerning the unsteady-state laminar flow of an incompressible, isothermal (newtonian) fluid in a cylindrical vessel spinning about its symmetry axis, say $z$, and inside which the liquid motion starts with an axial velocity component as well. 
Basic physical assumptions are that the pressure axial gradient keeps itself on its  hydrostatic value and that no radial velocity exists. In such a way the Navier-Stokes
PDEs become uncoupled and can be faced separately. We succeed in computing both the unsteady speed components, i.e. the axial $v_z$ and the circumferential $v_\theta$ as well, by means of infinite series expansions of Fourier-Bessel type under time exponential damping.  Following this, we also find the unsteady surfaces of dynamical equilibrium, the wall shear stress and the Stokes streamlines.

{\bf Keywords}: Navier-Stokes equations, unsteady flow, eigenfunctions, Fourier-Bessel  expansion.
\end{abstract}
\section{Introduction}
\subsection{On the birth of Navier-Stokes equations}

The Navier-Stokes equations are a non-linear PDE system ruling the motion of a fluid. In essence, they represent the balance between the rate of change of momentum of an element of fluid and the forces on it, as  Newton's second law does for a particle. Newton himself did not explain well the nature of the forces between the particles in a continuum, but started ({\it Principia}, 1687) the dynamics of a viscous fluid in an intuitive form, even if a definitive  theory would be constructed by some engineers in the first quarter of 19th century.

After Newton, the second law was applied by Euler to continuum mechanics since 1750, leading up to the publication of his seminal memoir \cite{euler} in 1757. In this memoir, he provided the system ruling a {\it frictionless fluid} motion, compressible or not, under an arbitrary set of external forces, while his predecessors had worked on incompressible ones with one (D. Bernoulli and J. Bernoulli) or two (D'Alembert) degrees of freedom.
But the birth of the Navier-Stokes equations came much later: it is linked to the birth of elasticity theory which, at the beginning of the 19th century, was an important asset to engineers looking for a sound theory of the beam bending. 
In 1827, C.L. Navier extended, see \cite{navier}, his previous theory to hydrodynamics and this led him to a new internal force in Euler's equations: a new $\mu-$force\footnote{$\mu$ means dynamical viscosity of the incompressible newtonian fluid.} which was generated by the non-uniformity of the motion of the fluid.
Many years later, the Navier-Stokes equations, as we now know them, were deduced from various physical hypotheses, for instance the stress  linearly related to the strain rate of the fluid, by G.G. Stokes, \cite{stokes}, who, reviewing the methods and hypotheses of Navier and other authors, presented a short rational approach to the equations of {\it viscous fluids}.
In fact fluids are classified on the rate at which they deform in response to an imposed shear stress: in newtonian fluids there is a linear relation between the shear stress and the strain rate, whereas in non-newtonian ones this relation is non-linear. 
The Navier-Stokes equations rule the motion of fluids in general, and are applicable to newtonian as well as non-newtonian fluids, to both laminar and turbulent flows of liquid/gas.

The main difficulty in solving them is their nonlinearity arising from the convective acceleration terms which, apart the unsteady ones, are all the left-hand side terms in \eqref{sis1}. There are no general analytical schemes for solving nonlinear PDE, and each problem must be considered individually. Unfortunately for most flow problems, fluid particles really do have acceleration as they move from one location to another in the flow field: thus,  the convective acceleration terms alone are usually not negligible. However, there are a few special cases, as in our paper, for which the convective acceleration vanishes, for instance due to the flow system geometry or to other cause. In these cases exact solutions, for an unsteady problem too, become possible. 
%%%%%%%%%%%%%%%%%%%%%%%%%%%%%%%%%%%%

%%%%%%%%%%%%%%%%%%%%%%%%%%%%%%%%%%%%%%

The word \lq\lq exact solution'' has a special meaning, denoting a simple, explicit  expression in finite terms either of elementary\footnote{See for instance \cite{knyazev}} or  well-known special functions. This is in contrast to an \lq\lq approximate solution'', which approximates a solution either in a numerical sense or in its asymptotic limit, for instance with a vanishingly small viscosity, and so on. The exact solutions are, essentially, a subset of  of the Navier-Stokes solutions which happen to have relatively simple mathematical expressions and are mostly due to a strongly symmetrical context. From the early 20$^{\rm th}$ century, asymptotic methods which extended the range of tractable problems after numerical methods were developed. Even though the emphasis of research in last twenty years changed from searching exact solutions to some approximate problems to finding approximate solutions to exact problems the exact solutions remain a valuable resource because they allow cross-checks of some numerical approaches and immediately convey more physical insight than all the numerical tables one could compute. This is specially true when the system is ruled by one or more parameters, so that a complete numerical tabulation would be neither practical nor clear. 

The functions used for our solutions are quite simple as consisting of infinite series of exponential and Bessel functions of integer order: by their analytic structure we are allowed to infer the main flow behaviors just before creating long computations.

\subsection{Outline of some contemporary literature}
There is not a rich collection of unsteady analytical Navier-Stokes solutions: let us provide here an up-to-date outline of them concerning a newtonian, incompressible fluid. We would then exclude some sound treatments like those of Ayub et alii \cite{hayat2006some} where our same aims are pursued but for fluids (those of magnetohydrodynamics) by a far different nature.

Our starting point could be the old, but authoritative, treatise due to Batchelor \cite{batchelor2000} where not more than a dozen pages are devoted to the unsteady unidirectional flow. He does not provide many details about the solutions mathematical generation and founds his treatments on the Stokes standard solution with the error function, trying to put back each unsteady 1-D problem to that one. The work of Drazin and Riley \cite{drazin} is less than a quarter of Batchelor's one, but the coverage of the unsteady topic is about half a book. Chapter 4 describes some unsteady flows bounded by plane boundaries (and then both Stokes's problems) considering in some detail the cases of an oscillating plate, the impulsive flows and the angled flat plate whose solution is expressed through the confluent Kummer hypergeometric function.
 Chapter 5 is devoted to the unsteady axisymmetric flow in pipes and particularly: pipes with a variable radius, impulsive flow, periodic flow. The pulsed flow, namely
with a periodic pressure gradient (also called Womersley’s flow) is also discussed.
 However it is not enough to believe that the authors always perform a complete analysis of the Navier-Stokes system: some basic solutions are taken as kernels and then adapted, making use of the continuity equation and eventually approaching to some ODE, being unsteady, but always 1-D problems.
 
The article of Knyazev \cite{knyazev} founds on the so called Karm\'{a}n family of the Navier-Stokes exact solutions. Some non-self-similar solutions are considered to solve the problem of an unsteady incompressible flow between two rotating disks, one of which moves along the common rotation axis. Interesting and detailed plots represent the solutions computed in terms of long but elementary functions.

We refer now to the fourth chapter of the book from Langlois and Deville \cite{langlois} where some similar unsteady problems are treated, following the alternative way of solving the Navier-Stokes PDE system and not the stream function. First of all they introduce the famous Stokes problems about the motion of a viscous fluid produced by a flat plate in a direction parallel to its plane. The motion of the plate may be either steady from a starting instant on (first Stokes problem: flow in a semi-infinite space), or periodic in time (second Stokes problem): since the plate oscillates with frequency $\cos(\omega t)$, a justified guessed solution can be $u(y,t) = f(y)\exp(i\omega t)$, being $i$ the imaginary unit. 
It is also treated the transient of the channel flow with a pulsatile pressure gradient or a Poiseuille flow with the pipe wall forced by a torsional oscillation, the angular speed changing as $\cos(nt)$.
The pulsating flow in a circular pipe has a pressure gradient that is oscillating in time and such a case is analyzed passing through the complex quantities.
A last good problem worthy to be cited: let us now imagine that a spherical $R$-bubble of inviscid gas is contained in an unlimited liquid. Suppose further that the pressure of the gas forming the bubble varies with time: the radius of the bubble will also change with time. Such a pulsating bubble will generate a velocity field within the liquid which in turn produces a stress field. At the end one is led to a second order non-linear differential equation, for which Langlois returns to his original articles.

In the recent book of Brenn \cite{brenn} the hydrodynamic problem is faced not directly through the velocity components but via the Stokes stream function PDE. The author collects some recent research papers. A first example is a spatially two-dimensional linear unsteady flow, with the flow velocity varying with the coordinate $y$, but not with $x$.
 Another case: flow fields along infinite structures without any geometrical elements with length scales, for instance along flat plates, without an imprinted flow time scale.  In every case, the form assumed for the velocity profile caused the nonlinear inertia terms in the Navier-Stokes equation either to vanish completely or to produce only a centrifugal force, easily balanced by a pressure gradient. It is then possible, or convenient, to perform a {\it linearization}. The exact solutions presented so far are by him called \lq\lq degenerate'': the linearization for cylindrical flows is solved by the approach of Tomotika. That for spherical flows leads to an equation for which the same approach leads to the Legendre functions of the first and second kind.

In some cases the unsteady problem leads to some nonlinear ODE as in Shapiro's paper \cite{shapiro} about an unsteady axisymmetric incompressible case with a decelerating backward stagnation-point flow with uniform injection or suction from a porous boundary (plate). The author arrives at a third order nonlinear ODE which after some work is transformed into a Riccati (tractable) equation.
 
To summarize, the exactly-solved unsteady problems concern few and really simple geometries, but different physical situations. The mathematical tools are: similarity variable transformation, variable separation, eigenfunctions expansion, so that several special functions are involved. It is less seen the employ of integral transforms of Fourier or Laplace, probably due to difficulties with the Bromwich contour integration.
 
\subsection{Aim of the work}\label{aimow}
Our system is consisting of a cylindrical vessel whose vertical axis, say $z,$ holding an incompressible liquid at rest is suddenly put in rotation with angular velocity $\bm{\omega}=\Omega \bm{k},$ where $\Omega \in \mathbb{R},$  starting from time $t=0$. The aim of this paper is the study of internal, laminar unsteady motion in both directions, that is axial and radial. As a matter of fact the liquid is loaded to  the steady loads of centrifugal and gravitational nature and, in addition, to a vertical outside pulse-to the start up only- which is responsible of the initial axial upward velocity $v_z^0 >0$.
Then we have $ {\bf v}=(v_r, v_\theta ,v_z).$ For the laminar incompressible viscous fluid, the Navier-Stokes equations in cylindrical coordinates $\left(r,\,\theta,\,z\right)$ are:
\begin{subequations}\label{sis1}
\begin{align}
&\rho \left[\dfrac{\partial v_r}{\partial t} +v_r \dfrac{\partial v_r}{\partial r}+ \dfrac{v_{\theta}}{r} \dfrac{\partial v_r}{\partial \theta}-\dfrac{v_\theta ^2}{r}+v_z \dfrac{\partial v_r}{\partial z} \right]= \nonumber\\
&=-
\dfrac{\partial P}{\partial r}+ \mu \left[\dfrac{\partial}{\partial r } \left( \dfrac{1}{r} \dfrac{\partial(r v_ r)}{\partial r} \right) + \dfrac{1}{r^2} \dfrac{\partial^2 v_r}{\partial \theta^2}-\dfrac{2}{r^2} \dfrac{\partial v_ \theta}{\partial \theta}
+ \dfrac{\partial^2 v_r}{\partial z^2}\right]+\rho g_r\label{radial}
\vspace{4mm} \\
&\rho \left[\dfrac{\partial v_\theta}{\partial t} +v_r \dfrac{\partial v_\theta}{\partial r}+ \dfrac{v_{\theta}}{r} \dfrac{\partial v_\theta}{\partial \theta}+\dfrac{v_r v_\theta }{r}+v_z \dfrac{\partial v_\theta}{\partial z} \right]=\nonumber \\
&=-
\dfrac{1}{r}\dfrac{\partial P}{\partial \theta}+ \mu \left[\dfrac{\partial}{\partial r } \left( \dfrac{1}{r} \dfrac{\partial(r v_\theta)}{\partial r} \right) + \dfrac{1}{r^2} \dfrac{\partial^2 v_\theta}{\partial \theta^2}+
\dfrac{2}{r^2} \dfrac{\partial v_ r}{\partial \theta}
+ \dfrac{\partial^2 v_\theta}{\partial z^2}\right]+\rho g_\theta\label{circumferential} \vspace{4mm} \\
&\rho \left[\dfrac{\partial v_z}{\partial t} +v_r \dfrac{\partial v_z}{\partial r}+ \dfrac{v_{\theta}}{r} \dfrac{\partial v_z}{\partial \theta}+v_z \dfrac{\partial v_z}{\partial z} \right]= \nonumber\\
&=-
\dfrac{\partial P}{\partial z}+ \mu \left[\dfrac{1}{r}\dfrac{\partial}{\partial r } \left( r \dfrac{\partial v_ z}{\partial r} \right) + \dfrac{1}{r^2} \dfrac{\partial^2 v_z}{\partial \theta^2}+ \dfrac{\partial^2 v_z}{\partial z^2}\right]+\rho g_z\label{axial}
\end{align} 
\end{subequations}
hereinafter called radial \eqref{radial}, circumferential \eqref{circumferential}, axial \eqref{axial}, ones, respectively.

Setting the non-vertical components of gravity to zero and the vertical one to $-g$, we assume that  {\bf the velocity radial component is always zero}, $v_r=0$ for any $r,\,\theta,\, z,\, t,$ thus $ {\bf v}=\left(0,\, v_\theta (r,t),\, v_z (r,t)\right).$ In such a case the circumferential and axial equations of Navier-Stokes system become uncoupled, so that they can be treated separately, as it is explained below.

%\begin{itemize}
\noindent $\bullet$
Equation in $ v_{z}$: we assume {\bf the axial pressure gradient does not differ from its hydrostatic asset}, namely: $\partial P/ \partial z=- \rho g$. In such a way we will find separately $v_{z}(r, t)$ as independent on the height  $z$. Such a velocity component must meet the upwards initial condition $ v_{z}(r, 0)=v_z^0 >0$  and in addition the {\it no-slip} one around the circular boundary: $ v_{z}(R, t)=0,$ for any $t>0$. 
Finally, when the transient is off, $v_{z}$ has to be extinguished, not being any physical reason for any vertical motion goes on within the time. 
%Inoltre, a transitorio estinto, essa non deve sopravvivere, dato che non esiste ragione fisica per cui si mantenga nel tempo un moto verticale.

\noindent $\bullet$ 
About the circumferential equation, assuming $v_\theta$ not depending on the height $z$, we will not do any further special assumption. 

\noindent $\bullet$
The radial equation becomes:
\[
\rho \dfrac{v_\theta ^2}{r}=\dfrac{\partial P}{\partial r}
\] 
then, after $v_\theta$, we will go on to integrate the pressure scalar field $P(r,z,t)$ whose axial gradient had been assumed to be known.
Notice that such a equation constitutes the nonlinearity of the Navier-Stokes  system, which has become an algebraic-differential: linear in its differential part but algebraically nonlinear. Such nonlinearity will make harder the field $P(r,z,t)$ computation.

\noindent $\bullet$ The computation of $v_\theta$ implies we can evaluate the unsteady equilibrium surfaces of liquid and the wall shear stress as well.
%\end{itemize}
\section{The axial sub-problem}

Let the $z$-velocity component  depend on $r$ and $t$ only: $v_z=v_z(r,t)$ and the $z$-transient flow does not affect the pressure axial gradient along such a direction:
\[
\dfrac{\partial P}{\partial z}=- \rho g.
\]
This does not imply that the unsteady overpressures are physically much lesser than the relevant static ones but only that axial pressure gradient is really ruled only by the weights and the heights, so that it changes little due to the small change in height.
With this assumption, the axial Navier-Stokes \eqref{axial} becomes linear:
\begin{equation}
 \dfrac{\partial v_z}{\partial t}=\frac{\nu}{r}\dfrac{\partial}{\partial r } \left( r \dfrac{\partial v_z}{\partial r} \right)
\label{ee7}
\end{equation}
to be solved under the conditions: 
\begin{subequations}\label{cond1}
\begin{align}
&v_z(0,t) \in \mathbb{R} \quad\text{finite velocity along the axis} \\
&v_z(R,t)=0 \quad\text{no-slip condition} \\
&v_z(r,0)=v_z^0 \quad\text{initial condition}
\end{align} 
\end{subequations}
Let us look for a solution to \eqref{ee7} by separation of variables: $v_z(r,t)=F(r)T(t).$  We get:
\begin{equation}
\begin{cases}
\dfrac{\dot{T}}{T}=-\eta^2 \\
\nu \left(\dfrac{F'}{rF}+\dfrac{F''}{F} \right)=-\eta^2,
\end{cases}
\end{equation}
being $\eta$ the (unknown) separation constant. The radial equation:
\begin{equation}
F''+\frac{F'}{r}+\frac{\eta^2}{\nu}F=0
\label{bessel}
\end{equation}
%la (\ref{bessel})%
is a special Bessel equation whose solution is:
\begin{equation*}
F(r)= C J_0 \left( \frac{\eta}{\sqrt{\nu}} r \right)+D Y_0 \left( \frac{\eta}{\sqrt{\nu}} r \right),
\end{equation*}
being $J_0$ the Bessel function of first kind and zero order and  $Y_0$ that of 
second kind and zero order.
\subsection{Initial and boundary conditions to  $\boldsymbol{v_z(r,t)}$}\label{contor}
Let us now provide details on the above relevant conditions.

%\begin{itemize}
\noindent $\bullet$ 
{\bf Finiteness}: due to the divergent behavior to $-\infty$ of $Y_0$ when its argument vanishes,  it shall be $D=0$, so that :
\begin{equation*}
F(r)= C J_0 \left( \frac{\eta}{\sqrt{\nu}}\, r \right).
\end{equation*}
Therefore:
\begin{equation}
v_z(r,t)= C J_0 \left( \frac{\eta}{\sqrt{\nu}} r \right)e^{-\eta^2t},
\label{sol}
\end{equation}
with $C$ and $\eta$ yet unknown.

\noindent $\bullet$
 {\bf Boundary condition}: the liquid velocities appearing in all previous formulae are all absolute: as adherent to the wall, the liquid shall move as the wall itself which is axially at rest, so that the condition along the circular boundary is: 
 \begin{equation*}
v_z(R,t)=0, \ t>0
\end{equation*}
 so that:
\[
J_0 \left( \frac{\eta}{\sqrt{\nu}} R \right)=0.
\]
Then:
\[
\frac{\eta_n}{\sqrt{\nu}} R= \beta_n \rightarrow \eta_n= \frac{\sqrt{\nu}}{R} \beta_n,
\]
where $\beta_n$  is the sequence of all roots of $J_0$. Putting in (\ref{sol}) we get: $$v_z(r,t)=C J_0 (\sigma_n r)e^{-\nu \sigma_n^2t},$$ 
where:
$$\sigma_n=\frac{\beta_n}{R}.$$
\noindent $\bullet$
{\bf Initial condition}:  we require that $v_z(r,0)=v_z^0 , \ v_z^0 \in \mathbb{R}$
where $v_z^0$ is the (e.g.) positive value of the initial upwards velocity intensity. We get:
$$v_z^0 =C\, J_0 (\sigma_n r).$$
%\end{itemize}

\begin{figure}[h]
\centering
\includegraphics[scale=0.38]{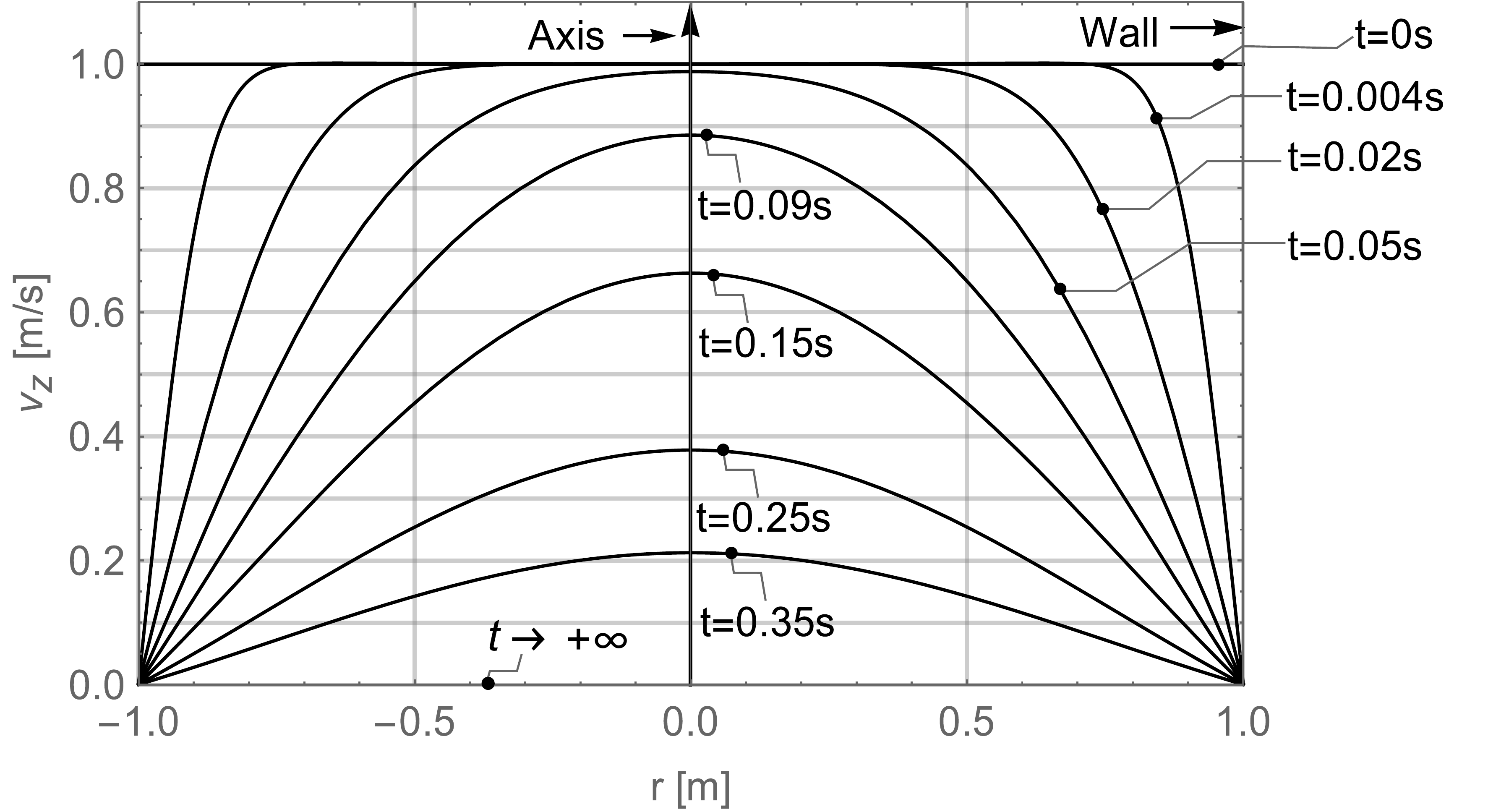}
\caption{Unsteady axial velocity with $v_z^0=1$, $R=1$, $\nu=1$}\label{figone}
\end{figure}

Generally such an initial condition will not be satisfied, but a superposition of infinite modes of $J_0$'s will be necessary: it will then be met not only by one but by infinite eigenfunctions:
$$v_z^0 =\sum_{n=1}^{+\infty}C_n J_0 (\sigma_n r)$$
each of them being connected to $J_0$ roots. The coefficients $C_{n}$ are yet unknown, but only one of them will survive. In fact, by multiplying to $r$ both sides of last one, after integrating between $r=0$ ed $r=1$, recalling the orthogonality properties of Bessel functions as expressed by the Lommel integrals, we get:
\[
C_n= v_z^0 \, \frac{\displaystyle{\int_0^1 J_0(\sigma_n r)r\,{\rm d}r}
}{\displaystyle{\int_0^1J_0^2(\sigma_nr)r\,{\rm d}r}}= v_z^0 \, \frac{\dfrac{1}{\sigma_n}J_1(\sigma_n)}{\dfrac{1}{2}J_1^2(\sigma_n)}.
\]
The axial unsteady flow is then ruled by:
\begin{equation}
v_z(r,t)=2v_z^0  \sum_{n=1}^{+\infty} \frac{J_0(\sigma_n r)}{\sigma_n  J_1 (\sigma_n)}e^{-\nu \sigma_n^2t}.
\label{soll}
\end{equation}
A large collection of 19th-century literature about the roots of Bessel functions is available: we will refer to McMahon formula \cite{macmahon}, quoted also in \cite{abramovitz} p. 371; specialized to our case, it becomes: 
\begin{equation*}
\begin{split}
\beta_n &= \pi (n-0.25)+\frac{1}{8  \pi (n-0.25)}-\frac{31}{384\pi^3 (n-0.25)^3}+ \\ &+ \frac{3779}{15360\pi ^5(n-0.25)^5}- \frac{6277237}{3440640\pi^7 (n-0.25)^7}+\ldots
\end{split}
\end{equation*}
We then computed the unsteady axial velocity: it can be used to investigate the flow changes produced by starting the rotation of the circular cylinder and the subsequent approach to a steady state.
The \figurename~\ref{figone} shows: the imposed uniform start value $v_{z}^0$, the fixed points of vanishing  $v_{z}$ at the round boundary, and all the declining profiles, whose maximum is gradually reduced with time, until complete extinction.
\begin{oss}
As a consequence, the volume flux across any section of the unsteady axial flow field,
$$
Q(t)=2\pi\int_0^R v_z(r,t) r\,{\rm d}r
$$
and the volume passed during the whole time
$$
\int_0^{+\infty} Q(t)\, {\rm d}t
$$
can be computed without difficulty by means of Bessel functions theorems.
\end{oss}
\section{The circumferential sub-problem}
Our treatment will be less short than the previous one due to the constraint between the circumferential velocity component and the radial pressure gradient. In addition, the PDE ruling the velocity is less simple. The relevant assumptions are:
%\begin{itemize}

\noindent $\bullet$ At a fixed $r$-value, the circumferential velocity component does not change  with the angle $\theta$ and the height $z:\,v_{\theta}= v_\theta(r,t), \quad\text{for any } \theta,\, z.$

\noindent $\bullet$ Pressure $P$ depends on time $t$ because we are dealing with an unsteady flow; on $r$ due to centrifugal force and, finally, on $z$ due to gravity. In addition there is no pressure gradient along the direction $\theta$, so that $P=P (r, z, t), \,\text{for any } \theta.$

\noindent $\bullet$  The radial component of velocity is assumed to be always zero $v_r=0$ for any $r,\,\theta,\, z,\, t.$

%\end{itemize}
In  radial and circumferential \eqref{circumferential}, and, \eqref{radial} Navier-Stokes equations, setting to zero the pressure circumferential gradient; as well as, the radial velocity and all the derivatives of  $v_\theta$  with respect to $\theta$ or $z$,  we get:
\begin{subequations}
\begin{align}
  & \rho \dfrac{v_\theta ^2}{r}=\dfrac{\partial P}{\partial r} \label{eq8}\\
  & \rho \dfrac{\partial v_\theta}{\partial t}= \mu\dfrac{\partial}{\partial r } \left( \dfrac{1}{r} \dfrac{\partial(r v_\theta)}{\partial r} \right) \label{eq4} \\
 & \dfrac{\partial P}{\partial z}=- \rho g
\label{eq3}
\end{align}
\end{subequations}

%\begin{subequations}
%\begin{align}
    %   r) & \ \ \rho \dfrac{v_\theta ^2}{r}=\dfrac{\partial P}{\partial r} \label{eq8} \\
      % \theta) & \ \  \rho \dfrac{\partial v_\theta}{\partial t}= \mu\dfrac{\partial}{\partial r } \left( \dfrac{1}{r} \dfrac{\partial(r v_\theta)}{\partial r} \right) \label{eq4} \\
     
%\end{align}
%\end{subequations}
\begin{oss}
Notice that the first of the equations above shows that the radial variation of pressure simply supplies the force necessary to keep the fluid elements moving in a circular path within the vessel.

\end{oss}

\begin{oss}
Using the above, the continuity equation:
\begin{equation*}
\frac{\partial \rho}{\partial t}+\frac{1}{r} \frac{\partial}{\partial r} (\rho r v_r)+ \frac{1}{r}\frac{\partial}{\partial \theta} (\rho v_\theta)+ \frac{\partial}{\partial z}(\rho v_z)=0
\end{equation*}
is automatically met for each time and for each point inside the fluid.
\end{oss}
\subsection{Circumferential sub-problem outline}
Ought to problem's linearity, let us arrange, see \cite{szymansky}, the function $v_\theta(r,t)$ in such a way:
\begin{equation}
v_\theta (r,t)=v_\theta^\infty(r)-\hat{v}(r,t)
\label{eq5}
\end{equation}
namely by adding to the steady state circumferential velocity distribution an unknown component $\hat{v}(r,t)$ depending on space and time.
The steady state solution $v_\theta^\infty$ can promptly be found: in fact, setting $\nu=\mu/\rho$ (kinematic viscosity), being $\partial_t v_\theta^\infty=0$, by(\ref{eq4}) we have:
\begin{equation*}
\dfrac{\partial}{\partial r } \left( \dfrac{1}{r} \dfrac{\partial(r v_\theta^\infty)}{\partial r} \right)=0.
\end{equation*}
A first integration leads to:
\begin{equation*}
\dfrac{1}{r} \dfrac{\partial(r v_\theta^\infty)}{\partial r}=K_1
\end{equation*}
Integrating again:
\begin{equation*}
v_\theta^\infty=\frac{K_2}{r}+\frac{K_1}{2} r.
\end{equation*}
Not being allowed any divergent behavior at finite, and then at $r=0$, then   $K_2=0$.

 Furthermore, due to the boundary condition at $r=R$ we have $v_\theta^\infty=\Omega R$, certainly it will be $K_1=2 \Omega$, so that: $v_\theta^\infty= \Omega r.$ Inserting in (\ref{eq5}) we have:
\begin{equation}
v_\theta (r,t)=\Omega r- \hat{v}(r,t).
\label{eq6}
\end{equation}
%Iniettando la (\ref{eq6}) nella (\ref{eq4}) si ha:
%\begin{equation*}
%\begin{split}
%- \frac{\partial \hat{v}}{\partial t} &= \nu \frac{\partial}{\partial r} \left[\frac{1}{r} \frac{\partial}{\partial r} \left(\Omega r^2-r \hat{v}\right) \right]= \nu \left(-\frac{\partial^2 \hat{v}}{\partial r^2}- \frac{r\dfrac{\partial \hat{v}}{\partial r}-\hat{v}}{r^2} \right)= \\
%&=\nu \left(-\frac{\partial^2 \hat{v}}{\partial r^2}-\frac{1}{r}\frac{\partial \hat{v}}{\partial r}+\frac{\hat{v}}{r^2} \right)
%\end{split}
%\end{equation*}
We then have the circumferential transient velocity equation:
\begin{equation}
\frac{\partial \hat{v} (r,t)}{\partial t}= \nu \left(\frac{\partial^2 \hat{v}(r,t)}{\partial r^2} + \frac{1}{r} \frac{\partial \hat{v}(r,t)}{\partial r}- \frac{\hat{v}(r,t)}{r^2} \right)
\label{eq7}
\end{equation}
which is linear and to be solved under \textbf{conditions}:
\begin{subequations}
\begin{align}
&\hat{v} (0,t) \in \mathbb{R}  \quad \text{along the $z$-axis the velocity shall be finite} \\
& \hat{v}(R,t)=0  \quad \text{adhesion to wall, usually referred as {\it no-slip condition}} \\
& \hat{v}(r,0)= \Omega r  \quad \text{functional type initial condition} 
\end{align}
\label{hhh}
\end{subequations}
At the same mathematical model, but as {\it external} motion, come \cite{lagerstrom} and \cite{drazin}.

\subsection{Integration}
In (\ref{eq7}) we put:
\begin{equation*}
r=e^x, \  \hat{v}(r,t) = u (x,t)
\end{equation*}
We have:
\begin{eqnarray*}
&& \frac{\partial \hat{v}}{\partial r}= \frac{\partial u}{\partial x} \frac{\partial x}{\partial r}= \frac{1}{r} \frac{\partial u}{\partial x} \\
&& \frac{\partial^2 \hat{v}}{\partial r^2}= \frac{\partial}{\partial r} \left( \frac{1}{r} \frac{\partial u}{\partial x}  \right)= - \frac{1}{r^2} \frac{\partial u}{\partial x}+\frac{1}{r} \frac{\partial^2 u}{\partial x^2} \frac{\partial x}{\partial r}=- \frac{1}{r^2} \frac{\partial u}{\partial x}+\frac{1}{r^2} \frac{\partial^2 u}{\partial x^2} 
\end{eqnarray*}
So that:
%\begin{equation*}
%\frac{\partial u}{\partial t}= \left[\frac{1}{r^2} \left( -  \frac{\partial u}{\partial x}+ \frac{\partial^2 u}{\partial x^2} \right)+\frac{1}{r^2} \frac{\partial u}{\partial x}- \frac{1}{r^2} u \right] \cdot \nu=\frac{\partial u }{\partial t}= \frac{\nu}{r^2} \left(\frac{\partial^2 u}{\partial x^2}-u\right)
%\end{equation*}
%E quindi:
\begin{equation*}
\frac{\partial u(x,t)}{\partial t}=\frac{\nu}{e^{2x}} \left[\frac{\partial^2 u(x,t)}{\partial x^2}- u(x,t) \right]
\label{equa2}
\end{equation*}
Let us look for a solution, by separation again:
 \begin{equation*}
u (x,t)=X(x)T(t)
\end{equation*} 
Marking by $\dot{}$ the time derivative and by $'$ the spatial one, we get: 
\begin{equation*}
X \dot{T}=\frac{\nu}{e^{2x}} \left( TX''- TX \right)
\end{equation*} 
%Dividendo per $XT$ si ottiene:
%\begin{equation*}
%\frac{\dot{T}}{T} =\frac{\nu}{e^{2x}} \left( \frac{X''}{X}-1  \right)
%\end{equation*}
It shall be: 
\begin{equation*}
\begin{cases}
\dfrac{\dot{T}}{T}=- \lambda^2 \\
\dfrac{\nu}{e^{2x}} \left( \dfrac{X''}{X}-1  \right)=-\lambda^2,
\end{cases}
\end{equation*}
being $\lambda$ the separation constants' set yet to be found. The first equation provides:
\begin{equation}
T(t)=e^{-\lambda^2 t} 
\end{equation}
Putting: $2x= \xi,$ the spatial equation becomes:
\begin{equation*}
X''(\xi)+ \left(\frac{\lambda^2}{4 \nu}e^{\xi}-\frac{1}{4}\right) X(\xi)=0
\end{equation*}
Making:
\begin{eqnarray}
\frac{\lambda^2}{4 \nu} = a >0, \label{cost1} \quad \frac{1}{4}=b >0,
\end{eqnarray}
then:
\begin{equation*}
X''(\xi)+ \left(a e^{\xi}-b \right)X(\xi)=0
\end{equation*}
whose general integral is:
\begin{equation}
X(\xi)=C J_{2 \sqrt{b}} \left(2 \sqrt{a} e^{\frac{\xi}{2}} \right) +D Y_{2 \sqrt{b}} \left( 2 \sqrt{a} e^{\frac{\xi}{2}} \right)
\label{eq2}
\end{equation}
(see \cite{Pol} p. 246). With $b=1/4$ from the second of \eqref{cost1}, then (\ref{eq2}) becomes:
\begin{equation}
X(\xi)=C J_{1} \left(2 \sqrt{a} e^{\frac{\xi}{2}} \right) +D Y_{1} \left( 2 \sqrt{a} e^{\frac{\xi}{2}} \right).
\label{JY}
\end{equation}

\subsection{Initial and boundary conditions for  $v_\theta (r,t)$}
They are:

%\begin{itemize}
\noindent $\bullet$
{\bf Finiteness}: it shall be $D=0$ and then:
\begin{equation*}
\hat{v}(r,t)=C J_1\left(\frac{\lambda}{\sqrt{\nu}} r\right)e^{-\lambda^2 t}.
\end{equation*}
\noindent $\bullet$
{\bf Boundary condition}: the no-slip condition with the wall implies a velocity  which shall for any $t>0$ be stuck on the value $\Omega R$ of the cylinder circumferential spinning velocity. In order to meet this, the unsteady contribution at $r=R$ shall be zero at any time. So that:
 $$ CJ_1 \left(\frac{\lambda}{\sqrt{\nu}} R \right)e^{-\lambda^2 {t}}=0.$$
And then:
\begin{equation*}
\frac{\lambda_n}{\sqrt{\nu}} R= \alpha_n
\end{equation*}
where $\alpha_n$ means the well-known sequence of roots of the Bessel  function $J_1$ (first kind and first order.
%\end{itemize}
By  (\ref{cost1}) we get the eigenvalues $\lambda_n$ of the separation constant:
\begin{equation*}
\lambda_n = \frac{\sqrt{\nu}}{R} \alpha_n
\end{equation*} 
Therefore:
\begin{equation}
\hat{v}(r,t)=C J_1 \left(b_n r\right)e^{- \nu b_n^2  t},
\label{margherita}
\end{equation}
where it has been put:
\[
b_n=\frac{\alpha_n}{R}.
\] 
% \begin{itemize}
\noindent $\bullet$
{\bf Initial condition}: 
Let us express the continuity of the junction between the velocity distributions in the sense that at the initial time the transient component must equate the steady component due to the fact that the overall velocity $v_\theta$ shall for any $r$ be zero at the start-up:
$$
v_\theta (r,0)=0 \quad\text{implies} \quad \hat{v}(r, 0)= v_\theta^\infty (r),
$$
this is the initial condition for $\ r \in [0,R]$. 
%\end{itemize}
The problem will be solved by expanding in a series of eigenfunctions the function $\Omega r$: the relevant coefficients $C_n$ have to be computed founding on the fact that the family of the eigenfunctions forms an orthonormal complete system:
\begin{equation*}
\sum_{n=1}^{+ \infty} C_{n} J_1 \left(b_n r  \right)= \Omega r.
\end{equation*}
The coefficients $C_{n}$, can be found by multiplying both sides of the above expansion by:
\begin{equation*}
J_1 (b_m r)\,r
\end{equation*}
having fixed a particular integer $m$ and integrating for $r$ from $0$ to $1$, so that:
\begin{equation*}
\int_0^1 \sum_{n=1}^{+ \infty} C_{n} J_1 ( b_n r )J_1 (b_m r )r {\rm d}r= \Omega \int_0^1 J_1 (b_m )r^2{\rm d}r.
\end{equation*}
Due to Bessel functions orthogonality and Lommel integrals, only the integral with $n=m$ will survive:
\begin{equation*}
C_m= \dfrac{\displaystyle{ \Omega \int_0^1 J_1 \left(b_m r  \right)r^2\, {\rm d}r}}{\displaystyle{\int_0^1 J_1^2 \left(b_m r  \right) r\,{\rm d}r}}
\end{equation*}
For the integral at numerator see formula 6.561/5 p. 683 and 672 of \cite{gradshteyn}. Then we found the eigenvalues discrete spectrum of $C_n$ for any $n \in \mathbb{N}$:
\[
C_n= \Omega \dfrac{\left(b_n \right)^{-1} J_2 \left(b_n \right)}{\dfrac{1}{2} J_2^2 \left(b_n \right) }= \frac{2 \Omega }{J_2 \left(b_n \right) b_n},
\]
being $J_2$ the first kind, order 2, Bessel function. 
So:
\begin{equation}
\hat{v}(r,t)= 2 \Omega \sum_{n=1}^{+\infty} \frac{J_1 (b_n r)}{b_n J_2(b_n)} e^{- \nu b_n^2 t}.
\label{bella}
\end{equation}
Therefore the overall circumferential velocity is given by:
\begin{equation}
v_\theta(r,t)=\Omega r-2 \Omega \sum_{n=1}^{+\infty} \frac{J_1 (b_n r)}{b_n J_2(b_n)} e^{- \nu b_n^2 t}
\label{mastra}
\end{equation}
 Again, as previously \eqref{contor} for roots $\beta_n$ of $J_0$, in order to compute all roots $\alpha_n$ of  $J_1$ function, we will use the McMahon formula which, for our case becomes:
 \begin{equation*}
 \begin{split}
\alpha_n =& \pi (0.25 + n)-\frac{3}{8  \pi (0.25 + n)}+\frac{3}{128  \pi^3 (0.25 + n)^3}-
 \\
& - \frac{1179}{5120  \pi^5 (0.25 + n)^5}+\frac{1951209}{1146880  \pi^7 (0.25 + n)^7}-\cdots
\end{split}
 \end{equation*}
being $n=1,\dots,+\infty$ the marker of the $n$-th zero of $J_1$. 
\subsection{ A profile analysis of ${v_\theta}$}
For all next computations, we took (SI units): $\Omega=1,\,R=1,\,\nu=1.$
The transient component $\hat{v}(r,t)$ is depending exponentially on time and in a Bessel way on the radius. 
\begin{figure}[h]
\centering
\includegraphics[scale=0.3]{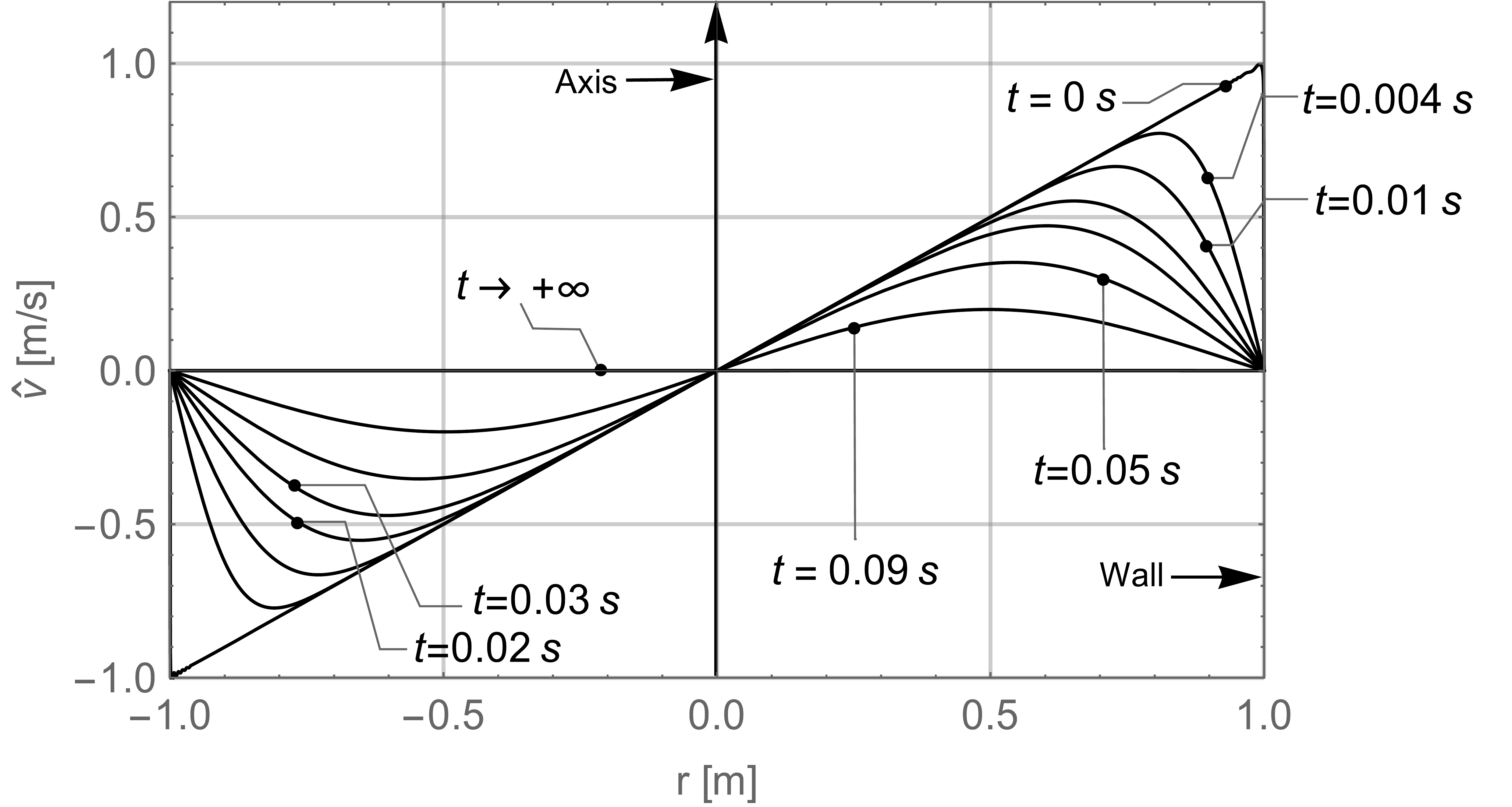}
\caption{Circumferential velocity unsteady response $\hat{v}(r,t)$: radial profiles.}
\label{figtwo}
\end{figure}

\begin{figure}[h]
\centering
\includegraphics[scale=0.30]{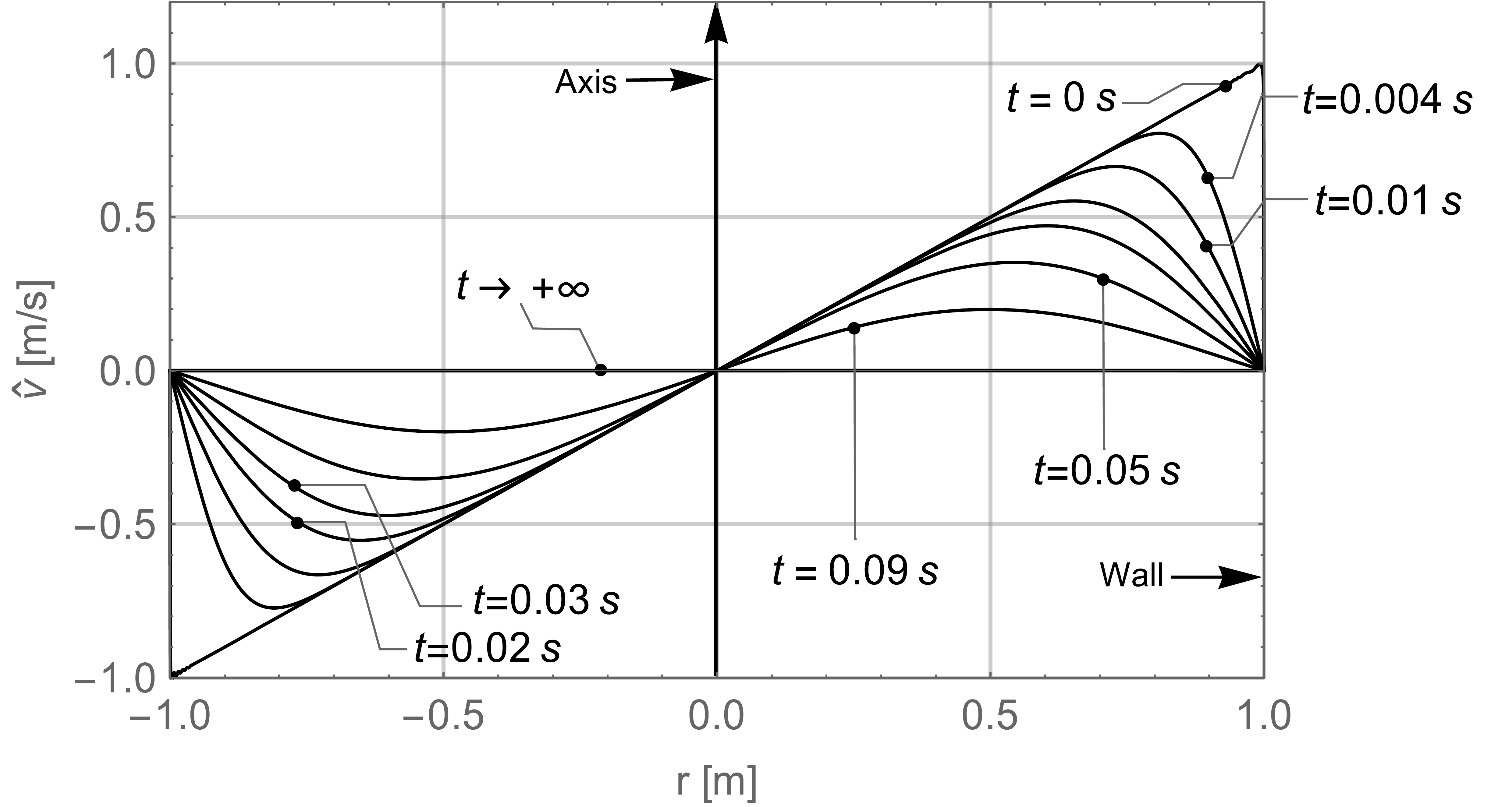}
\caption{Circumferential overall velocity: radial profiles, $v_\theta(r,t)$.}\label{figthree}
\end{figure}

At the start-up, being $\hat{v}(r,0)=v_\theta(r),$ the transient begins with a linear plot. Notice that we used really thousands terms in (\ref{bella}) for earning a correct representation of the linear profile in terms of a Fourier-Bessel expansion. Declining the transient with time, the initial linear plot will twist in the curves shown in the \figurename~\ref{figtwo}, up to its extinction. Furthermore notice that, by setting to zero, for a fixed time $\overline{t}$, the derivative with respect to $r$ of (\ref{bella}), we get an expression like:
$$
\sum_{n=1}^{+ \infty} \frac{e^{- \nu b_n^2 \overline{t}}}{J_2(b_n)} \cdot \frac{{\rm d} J_1}{{\rm d}r}(b_n \overline{r})=0,
$$
being $\overline{r}$ the $r>0$ value maximizing $\hat{v}(r)$. 
All the performed simulations showed that, for increasing time, the $\overline{r}$ sequence is going down: this is due to the exponential decay-law which rules the ${\rm d}J_1/{\rm d}r$ to become zero at decreasing $r$ values. When time increases, the exponential factor loses its weight and then the maximum of $\hat{v}$ does not change its position any more, see \figurename~\ref{figtwo}. The overall velocity $v_\theta$ profile is then shown by \figurename~\ref{figthree}.
\subsection{The pressure field}\label{pressuref}
By (\ref{eq8}) :
\begin{equation}
P(r,z,t)=\rho \int_0^r \frac{1}{\psi} v_\theta^2(
\psi,t){\rm d}\psi+f(z,t)+\beta
\label{alfa}
\end{equation}
reminding the pressure axial gradient to be always given by $-\rho g$, we get:
\begin{equation*}
P(r,z,t)=\rho \int_0^r \frac{1}{\psi} v_\theta^2(\psi,t){\rm d}\psi-\rho g z+ \beta.
\end{equation*}
In order to compute $\beta$ we set the condition: $P(0,0,t)=P_0$ being $P_0$ a given whichever value. Then $\beta=P_0$ and consequently the pressure field is univocally determined as:
\begin{equation}
P(r,z,t)=\rho \int_0^r \frac{1}{\psi} v_\theta^2(\psi,t){\rm d}\psi-\rho g z+ P_0
\label{eq11}
\end{equation}
and injecting there the $v_\theta$ law:
\begin{equation}
P(r,z,t)=P_0-\rho g z + \rho \Omega^2 \int_0^r \psi {\rm d}\psi+\rho \int_0^r \frac{\hat{v}^2(\psi,t)}{\psi}{\rm d}\psi-2 \rho \Omega \int_0^r \hat{v}(\psi,t){\rm d}\psi.
\label{pressione}
\end{equation}
So in the full expression of the overall  pressure, to the hydrostatic component two more dynamic components are added due to $ \hat{v} $ and then they decay with time. Notice that the last integral can be computed by adding infinite integrals of  type:
\begin{equation*}
\int_0^a J_1 (x){\rm d}x=1-J_0(a), 
\end{equation*}
where we invoked formula 6.511-7 page 660 \cite{gradshteyn}. The only one computation to be performed numerically is then the penultimate integral in \eqref{pressione}. 
\begin{oss} After the unsteady state has passed, the steady velocity is $v_\theta^\infty=\Omega r$ and, then, inserting in (\ref{eq11}) we have:
\[
P(r,z,\infty)=\frac{\rho \Omega^2}{2}r^2-\rho gz+P_0,
\] 
namely the well-known steady pressure distribution of centrifugal-gravitational nature.
\end{oss}

\subsection{The liquid free surface $z(r,t)$}
Let us put $P(r,z,t)=P_0$ in (\ref{pressione}): it means that we are looking for the space locus of the liquid points for which the internal pressure equates the atmospheric one, i.e. how the liquid free surface is shaped in space and time. We begin by doing a first simplification, taking into account the effect of the simultaneous transient  and stationary terms of $v_\theta$, which leads us to write (\ref{pressione}) as:
\begin{equation}
z(r,t)= \frac{1}{g} \left[\Omega^2 \frac{r^2}{2}-2 \Omega \int_0^r \hat{v}(\psi,t){\rm d}\psi \right].
\label{zfree}
\end{equation}

\begin{figure}[h]
\centering
\includegraphics[scale=0.39]{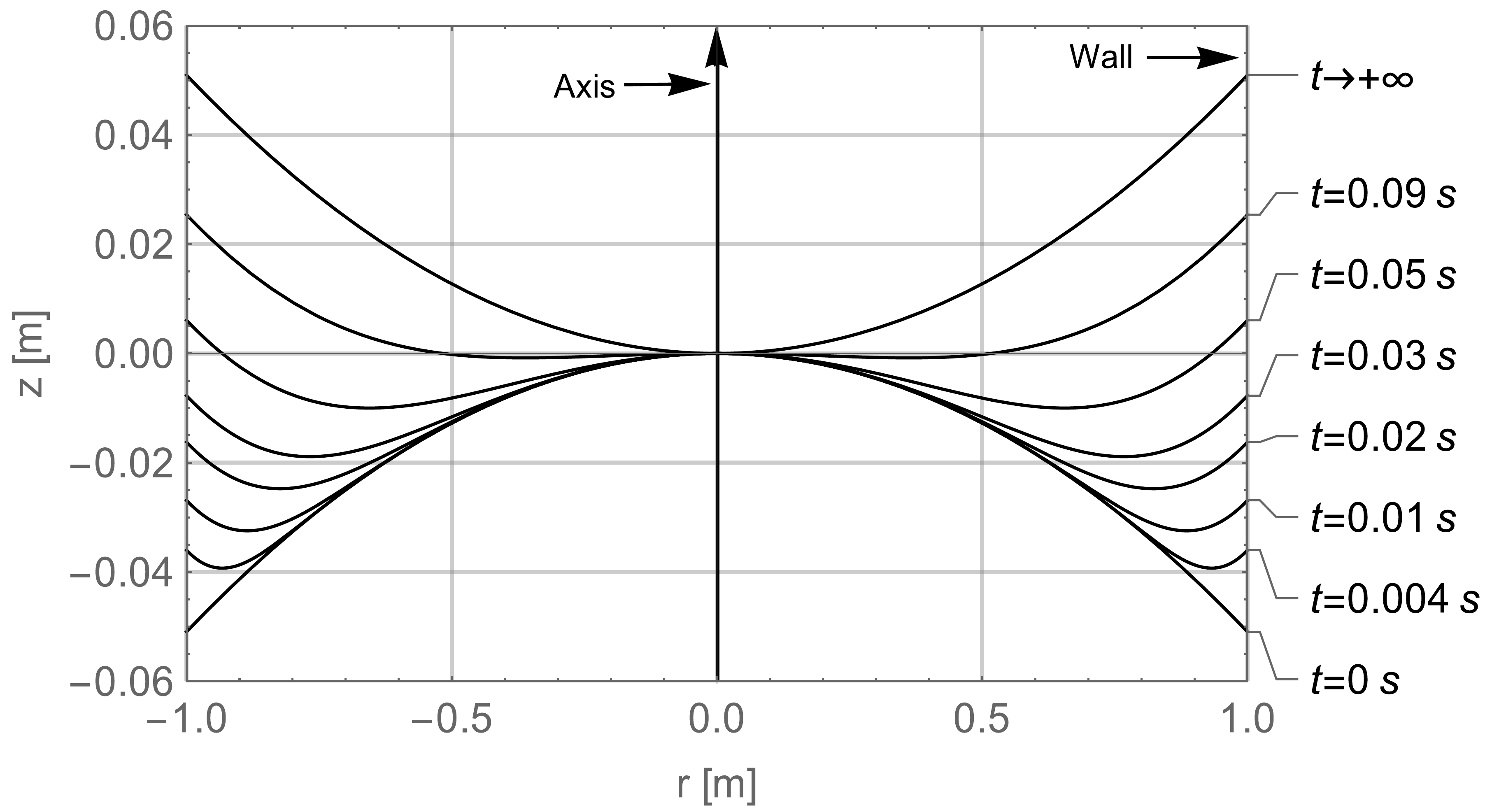}
\caption{The liquid free surface $z(r,t)$: transient response profiles. First approximation, $\frac{\hat{v}^2(r,t)}{r}={\rm o}(1)$.}\label{figfour}
\end{figure}
It has been neglected the penultimate term of  formula  (\ref{pressione}) as subjected to a more intense drop during time. The \figurename~\ref{figfour} shows that immediately after the start-up, the intersection of the liquid round surface with any plane holding the axis of rotation, provides an array of initially concave lines. In an instant between $0 \ s$ and $0.004 \ s$ (that can be calculated numerically), there is a change of curvature. Next, the curves become convex till when, extinguished almost the transient ($t >0.09 \ s$), the free surface takes the known shape of round paraboloid.
The surface description does not seem acceptable for the first time values, when the contributions to the overall solution, brought by the integral quadratic which we neglected in the first time, are not yet negligible.

Mind that for $t=0$ we have $v_\theta=0$ and then from (\ref{eq11}) we get $z(r,t)=0$; the $z(r)$ profile is then initially flat for any $r$, as expected. The height $z(r,t)$ is then given by:
\[
z(r,t)= \frac{1}{g} \left[\Omega^2 \frac{r^2}{2}+\int_0^r \frac{\hat{v}^2(\psi,t)}{\psi}{\rm d}\psi-2 \Omega \int_0^r \hat{v}(\psi,t){\rm d}\psi \right].
\]
By the initial condition $\hat{v}(r,0)=\Omega r$, we take that:
\[
\frac{\hat{v}^2(r,t)}{r}=\Omega \frac{\hat{v}^2(r,t)}{\hat{v}(r,0)}
\]
therefore, since in the following we need to study the $t$-dependence of the following integral we inttoduce:
\begin{equation} 
I(t,r) = \int_0^r \frac{\hat{v}^2(\psi,t)}{\psi} {\rm d}\psi= \Omega\int_0^r  \frac{\hat{v}^2(\psi,t)}{\hat{v}(\psi,0)}{\rm d}r.
\label{prova}
\end{equation}
We now establish a map of the time evolution of the integral $I(t)$ which is plotted by a different $r$-curve for each $t$. At start-up by (\ref{prova}) we get:
\begin{equation*}
I(0,r)=\Omega \int_0^r \hat{v}(\psi,t){\rm d}\psi=\Omega^2 \int_0^r \psi{\rm d}\psi= \frac{\Omega^2}{2}r^2. 
\end{equation*}
In a next time, say $\overline{t}=0+ \varepsilon$ with $\varepsilon \rightarrow 0^+$, we will get:
\begin{equation*}
I({\varepsilon},r)= \Omega\int_0^r \frac{\hat{v}^2(\psi,\varepsilon)}{\hat{v}(\psi,0)}{\rm d}r< I(0)=\frac{\Omega^2}{2}r^2,
\end{equation*}
being $\hat{v}(r,\varepsilon)<\hat{v}(r,0)$ as shown by Figure \ref{figtwo}.
Therefore for each next time, the curves plotting the integral $I(t,r)$ will lay below the parabola graph of the integral $I(0,r)$. For $t \rightarrow + \infty$ clearly we have $I(\infty,r)=0$.
\begin{figure}[h]
\centering
\includegraphics[scale=0.35]{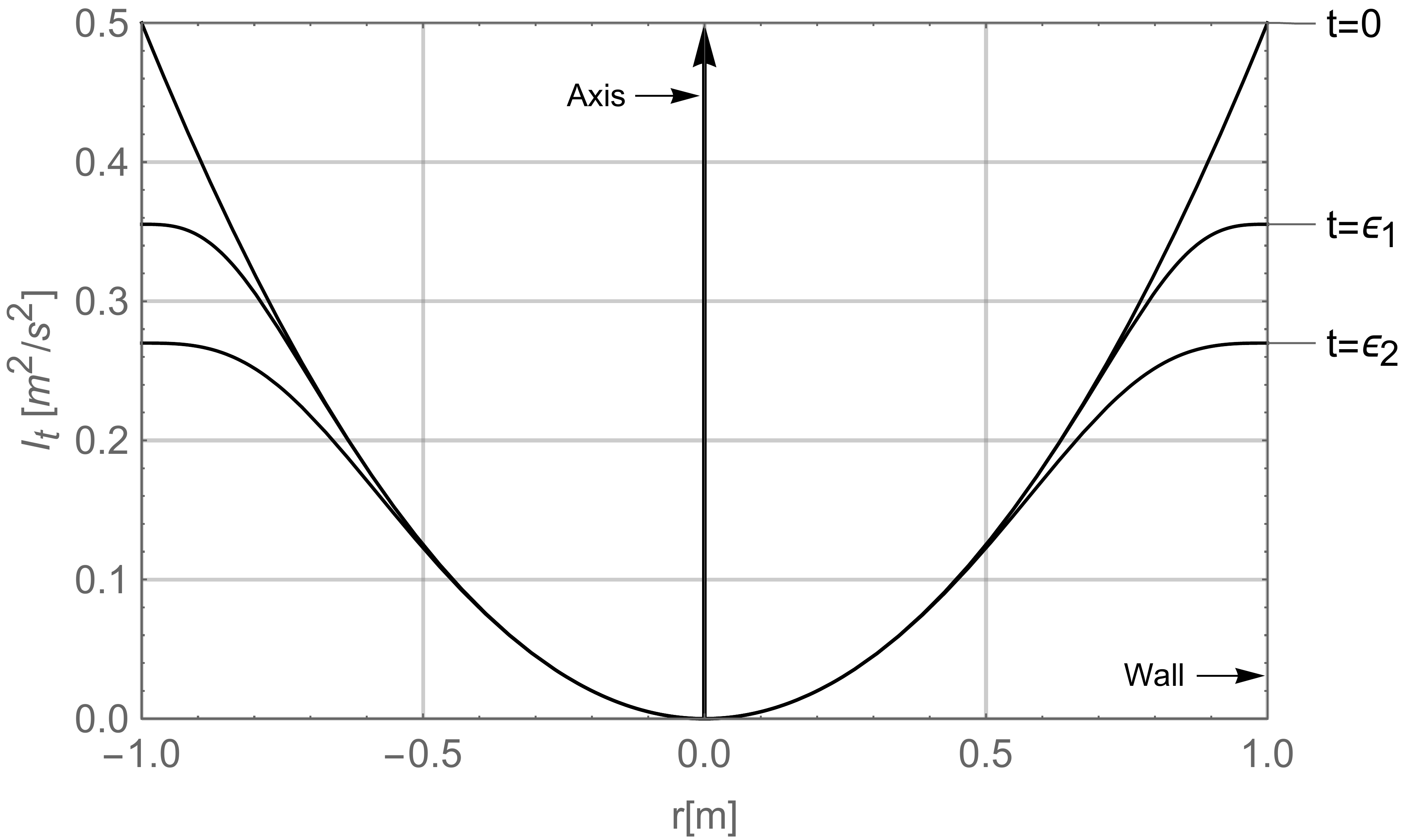}
\caption{Time evolution of integrals $I(t)$, for some $t$, with $\varepsilon_2>\varepsilon_1$}
\end{figure}
Through a numerical computation of the integral of $\frac{\hat{v}^2(r,t)}{r}$, the overall solution of the free round surfaces is shown by \figurename~\ref{figsix}: 
\begin{figure}[h]
\centering
\includegraphics[scale=0.38]{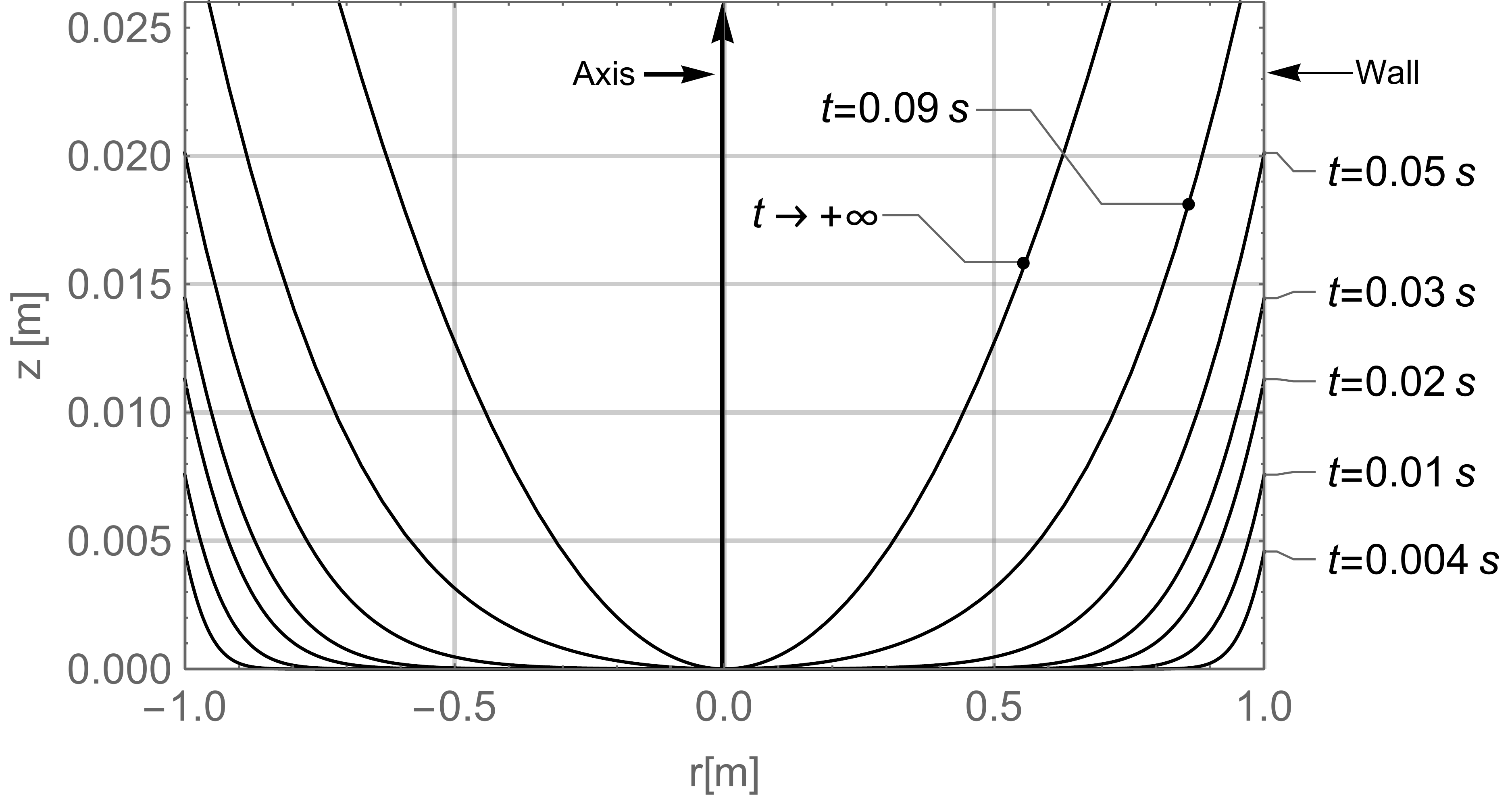}
\caption{The unsteady free surfaces profiles: full evaluation.}\label{figsix}
\end{figure}

\subsubsection{About the  $\boldsymbol{r}$-integral of $\boldsymbol{\hat{v}^2(r,t)/r}$}
The mentioned integral has been already evaluated numerically: nevertheless we wish to give hereinafter a formal treatment of it founded on the $_2{\rm F}_1$ hypergeometric function. We have to integrate with respect to $r$:
\begin{equation}\label{equired}
I(t,r)=\int_0^r\frac{\hat{v}^2(\psi,t)}{\psi}{\rm d}\psi,
\end{equation}
where
\begin{equation}
\hat{v}(r,t)=2\Omega \sum_{n=1}^{+\infty}\frac{J_1(b_nr)}{b_nJ_2(b_n)}e^{-\nu b_n^2t}=2\Omega \sum_{n=0}^{+\infty}\frac{J_1(b_{n+1}r)}{b_{n+1}J_2(b_{n+1})}e^{-\nu b_{n+1}^2t}
\end{equation}
Let us apply the series product theorem (Cauchy):
\begin{equation}
\sum_{n=0}^{+\infty}a_n \cdot \sum_{n=0}^{+\infty}b_n= \sum_{n=0}^{+\infty}\sum_{k=0}^n a_kb_{n-k}
\end{equation}
So that:
\begin{equation}
\frac{\hat{v}^2(r,t)}{r}=4\Omega^2 \sum_{n=0}^{+\infty} \sum_{k=0}^n \frac{e^{-\nu t \left(b_{k+1}^2+b_{n-k+1}^2\right)}}{b_{k+1}b_{n-k+1}J_2(b_{k+1})J_2(b_{n-k+1})}\, \frac{J_1(b_{k+1} r)J_1(b_{n-k+1}r)}{r}
\end{equation}
The term by term integration with respect to $r$ in order to compute \eqref{equired} concerns the term:
\begin{equation}
\int_0^r \frac{J_1(b_{k+1} \psi)J_1(b_{n-k+1}\psi)}{\psi}{\rm d}\psi
\label{maledetto}
\end{equation}
The product of Bessel $J$-functions of different orders and arguments multiplied by a power of the variable can be integrated by means of formula 27 at page 259 of \cite{luke}:
\begin{equation}\label{lukkk}
\begin{split}
\int_0^z t^{\rho}J_{\mu}(at)J_{\nu}(bt){\rm d}t&=\frac{(az/2)^{\mu}(bz/2)^{\nu}z^{\rho+1}}{\Gamma(\mu+1)\Gamma(\nu+1)}\times\\
&\times\sum_{k=0}^{+\infty} \frac{(-1)^k (az/2)^{2k}}{k!(\mu+\nu+\rho+2k+1)(\mu+1)_k}\,\,_2{\rm F}_1 \left( \left. 
\begin{array}{c}
-k,-\mu-k \\[2mm]
\nu+1
\end{array}
\right| \frac{b^2}{a^2}\right),
\end{split}
\end{equation}
where it shall be ${\rm Re}(\mu+\nu+\rho)>-1$, which in our case is fulfilled. Moreover, we do not have any problem of convergence for the hypergeometric series even if the ratio $b/a$ is not less than 1 due to the sequence of roots ruled by McMahon: nevertheless the $_2{\rm F}_1$ looses, the character of series terminating in a polynomial for being integer and negative both parameters $-k$ and $-\mu-k$ of its definition. Based on \eqref{lukkk}, the integral  \eqref{maledetto} will be given by:
\begin{equation}
\frac{b_{k+1}b_{n-k+1}r^2}{8}\sum_{m=0}^{+\infty}\frac{(-1)^m (b_{k+1}r/2)^{2m}}{m!(m+1)(2)_m}  \, _2{\rm F}_1\left( \left. 
\begin{array}{c}
-m,-1-m \\[2mm]
2
\end{array}
\right| \frac{b_{n-k+1}^2}{b_{k+1}^2}\right).
\end{equation}
The final expression of integral \eqref{equired} is then:
\begin{equation}\tag{\ref{equired}'} \label{equired'}
\begin{split}
I(t,r)&=\frac{(\Omega r)^2}{2} \sum_{\substack{n=0} }^{+\infty} \sum_{k=0}^n \frac{e^{-\nu t \left(b_{k+1}^2+b_{n-k+1}^2\right)}}{J_2(b_{k+1})J_2(b_{n-k+1})}\times\\
&\times\sum_{m=0}^{+\infty} \frac{(-1)^m (b_{k+1}r/2)^{2m}}{m!(m+1)(2)_m}\, _2{\rm F}_1\left( \left. 
\begin{array}{c}
-m,-1-m \\[2mm]
2
\end{array}
\right| \frac{b_{n-k+1}^2}{b_{k+1}^2}\right).
\end{split}
\end{equation} 
The practical employ of the formula \eqref{equired'} would probably require skills of program optimization in order to reduce conveniently the machine running time.

\section{Fluid motion representations}
\subsection{Streamlines}
 For a rotationally symmetric flow, we have: 
\begin{equation*}
\begin{cases}
x=r \cos(\theta) \\
y=r \sin(\theta)\\
z=z
\end{cases}, \
\begin{cases}
\textbf{i}=\cos(\theta)\textbf{e}_r-\sin(\theta)\textbf{e}_\theta \\
\textbf{j}=\sin(\theta) \textbf{e}_r+\cos(\theta)\textbf{e}_\theta\\
\textbf{k}=\textbf{k}
\end{cases}
\implies {\rm d}\textbf{s}= {\rm d}r \textbf{e}_r +r {\rm d}\theta \textbf{e}_\theta + {\rm d}z \textbf{k}
\end{equation*}
By the streamlines definition:
\begin{equation*}
\textbf{v} \wedge {\rm d} \textbf{s}=\textbf{0} \rightarrow 	\det\left( \begin{matrix}
\textbf{e}_r & \textbf{e}_\theta & \textbf{k} \\
v_r & v_\theta & v_z \\
{\rm d}r & r {\rm d} \theta & {\rm d}z 
\end{matrix}\right)= \left( \begin{matrix}
0 \\ 0 \\ 0
\end{matrix} \right)
\end{equation*}
In our specific case, given any instant $t_0$ (after it has to be considered as a variable parameter $\mathbb{R^+}$), we will get:
\begin{subequations}\label{e123}
\begin{align}
& v_ \theta (r,t_0) {\rm d}z = r  {v}_z {\rm d} \theta \label{e3} \\
& {v}_z {\rm d}r=0 \label{e1} \\
& v_\theta(r,t_0){\rm d}r=0 \label{e2}
\end{align}
\end{subequations}
By (\ref{e1}) or (\ref{e2}) we soon infer it shall be $r=r_0={\rm constant}$ so that, inserting in (\ref{e3}) and by integration with respect to $z$, we get a double infinity of streamlines which are helixes: 
\begin{equation}
\theta(r_0, t_0; z)= \dfrac{v_\theta(r_0,t_0)}{r_0 {v}_z (r_0,t_0)} z+\kappa_0,
\label{e4}
\end{equation}
being $z \in[0,H] ,\,\theta\in[0,2 \pi],\,\kappa_0=\theta(r_0,t_0, 0).$

\subsection{Shear stress  $\boldsymbol{\tau_{r,\theta}}$ at the wall}\label{shear}
Any real fluids (liquids and gases included) moving along solid boundary, will undergo a shear stress exerted on that boundary. The no-slip condition dictates that the velocity of the fluid relative to the boundary is zero. But at some height from the boundary the flow velocity must equate that of the unperturbed fluid. The region between these two points is aptly named the boundary layer. For all Newtonian fluids in laminar flow the shear stress is proportional to the strain rate in the fluid, the viscosity being the constant of proportionality.
Knowing the velocity field, we may derive the shear stress in the fluid due to the
spinning wall with the no-slip condition on its surface. 
As a consequence, at the wall of spinning vessel there is a shear stress $\tau_{r, \theta}$ directed circumferentially, $\theta$ being $r$ the direction orthogonal to the stressed area. Inserting $v_r=0$ in formula (D) of \cite{fdt} we get:
\begin{equation}
\tau_{r, \theta}=- \mu r  \frac{\partial}{\partial r} \left( \frac{v_\theta (r,t)}{ r}\right)
\label{taglio}
\end{equation} 
Notice that by inserting the  (\ref{mastra}) in  (\ref{taglio}), the wall shear stress, as expected, can be determined only by the transient component of the velocity $\hat{v}(r,t)$. By focusing on a particular  instant of time, we get the law of the shear stress in the fluid at different radial $r$ values. Repeating the process for several further instants of time, we will describe the shear stress with time. When evaluating the stress for $t \rightarrow 0$, such as physics suggests, $\tau_{r,\theta}=0$; this can be explained in two different ways.      

The first consists of saying that for $t \rightarrow 0$ one shall have $\hat{v}(r,t)=\Omega r$ (initial condition previously set) and then, injecting it in (\ref{taglio}) it is found $\tau_{r,\theta}=0$. 

The second one consists of performing derivatives of $\hat{v}(r,t)$, and injecting directly in (\ref{taglio}); a plot of the relevant curve for small times will show oscillations. But they are without any possible physical meaning and only due to the number of the terms chosen in order to represent $\hat{v}(r,t)$. As a matter of fact the number of terms which well fits a function, will not be able to do the same with its derivative. Accordingly, several computations and plots, here not attached, have been performed by making use of a lot of terms of the series for $\hat{v}$: and we checked that as their number increases, the pseudo-oscillation decay until to zero. After a threshold value $\overline{t}$, shear stress takes as follows:
\begin{figure}[h]
\centering
\includegraphics[scale=0.38]{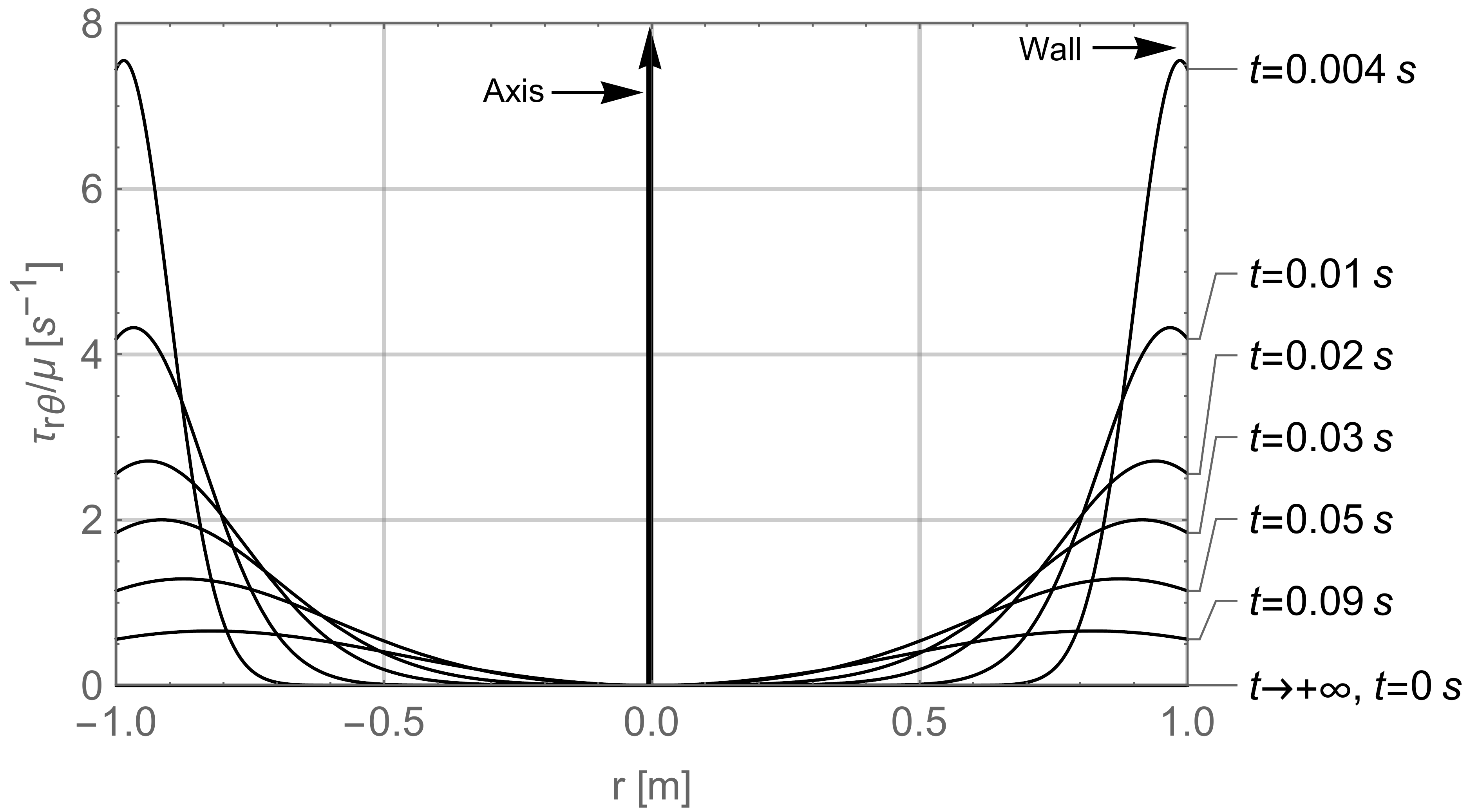}
\caption{The shear stress unsteady response: radial profiles.}\label{figseven}
\end{figure}
Because of the onset of a transient circumferential component of velocity, this produces a darting to the shear stress for any $r$ having its maximum at the wall. 
The family of the stress profiles shown in \figurename~\ref{figseven} is dying quickly and drives to equilibrium. After the circumferential stress has been computed as a function of  $r$ and time and it is specialized to $r=R$ and then by multiplying by $2 \pi R \cdot 1$, we get the magnitude of the circumferential shear stress  for unit height of the vessel.

\section{Conclusions}
The detection of Navier-Stokes analytical unsteady solutions has been rather trendy for the last 200 years, such that a (non-exhaustive) outline of the most recent literature (books  and papers) has been provided. Each possible analytical approach requires some physical assumption for passing to a tractable PDE system: our basic ones are, see our subsection \ref{aimow}, that the pressure axial gradient keeps itself on its hydrostatic value and no radial velocity arises.

We provided an integration of a polar coordinates Navier-Stokes equation system for an unsteady-state laminar flow of incompressible isothermal (newtonian) fluid in a cylinder spinning about its axis, and inside which the liquid starts with an axial velocity component as well.

The axial problem PDE \eqref{ee7} has been solved under conditions \eqref{cond1}, so that the relevant solution is given by formula \eqref{soll}, plotted at \figurename~\ref{figone}. The circumferential problem PDE is \eqref{eq7} to be solved under conditions \eqref{hhh}, so that the relevant overall solution is given by formula \eqref{mastra}, plotted at \figurename~\ref{figthree}.
The pressure field is then computed according to \eqref{pressione} liquid free surface is found through \eqref{zfree}, according to analysis in subsection \ref{pressuref} and \figurename~\ref{figsix}. Finally, \figurename~\ref{figseven} summarizes all the analyses concerning the shear stress unsteady response, discussed in detail at our subsection \ref{shear}.

%\bibliographystyle{elsarticle-num} 
%\bibliography{NavierStokes}

\end{document}